\documentclass[%
 aip,
 amsmath,amssymb,
 reprint,%
]{revtex4-2}

\usepackage{graphicx}% Include figure files
\usepackage{dcolumn}% Align table columns on decimal point
\usepackage{bm}% bold math
\usepackage{mathtools}
\usepackage[utf8]{inputenc}
\usepackage[T1]{fontenc}
\usepackage{mathptmx}
\usepackage{etoolbox}

\usepackage{xcolor}

\usepackage{braket}
\usepackage{hyperref}

\makeatletter
\def\@email#1#2{%
 \endgroup
 \patchcmd{\titleblock@produce}
  {\frontmatter@RRAPformat}
  {\frontmatter@RRAPformat{\produce@RRAP{*#1\href{mailto:#2}{#2}}}\frontmatter@RRAPformat}
  {}{}
}%
\makeatother
\begin{document}

\preprint{AIP/123-QED}

\title{Satellite radio detection via dual-microwave Rydberg spectroscopy}
%
% Force line breaks with \\
\author{Peter K Elgee}
 
\author{Joshua C Hill}%
% \email{Second.Author@institution.edu.}
\author{Kermit-James E Leblanc}

\author{Gabriel D Ko}

\author{Paul D Kunz}

\author{David H Meyer}

\author{Kevin C Cox}

\affiliation{ 
DEVCOM Army Research Laboratory, Adelphi MD 20783, USA%\\This line break forced with \textbackslash\textbackslash
}%

\email{peter.k.elgee.ctr@army.mil}

\date{\today}% It is always \today, today,

\begin{abstract}

Rydberg electric field sensors exploit the large number of Rydberg resonances to provide sensitivity over a broad range of the electromagnetic spectrum.
However, due to the difficulty of accessing resonant Rydberg states at ultra-high frequency (UHF) and below, ubiquitous bands in the world's current wireless communications infrastructure, they currently fall short in sensitivity in this range.
We present a resonant Rydberg electric field sensor operating in the UHF band using a dual-optical dual-microwave spectroscopy scheme.
Adding an additional microwave photon allows us to access transitions between Rydberg states with higher angular momentum ($L = 3 \rightarrow 4$), which have lower resonant frequencies than transitions typically used in Rydberg sensors.
We discuss the applicability of this type of sensor across the UHF band and below, and measure the resonant sensitivity of our system at 2.3~GHz to be 70(5)~$\mu$Vm$^{-1}\text{Hz}^{-1/2}$, 50 times better than the measured sensitivity with a far off-resonant probing scheme at this frequency.
We also show the effectiveness of this sensing scheme by measuring Sirius XM satellite radio (2.320 - 2.345~GHz) received outside the laboratory and rebroadcast onto the atoms.

\end{abstract}

\maketitle

Detecting radio frequency (rf) electromagnetic waves with frequencies between 30\,MHz and 3\,GHz is a ubiquitous and crucial task for many commercial applications including television, Wi-Fi, and cellular phones. Quantum electric field sensors based on Rydberg atoms have received considerable attention for such detection. A multitude of accessible discrete atomic resonances between Rydberg states with large electric dipole moments enable sensitivity to fields with frequencies spanning quasi-DC \cite{Jau2020} to THz \cite{Downes2020}. Simultaneous wideband detection is possible \cite{Meyer2023} in a single vapor-cell device and sensitivities down to ~$\mu$Vm$^{-1}\textrm{Hz}^{-1/2}$ and below have been demonstrated \cite{Jing2020, Parniak2023}.

Most contemporary Rydberg electric field sensors use two-photon electromagnetically induced transparency (EIT) to establish a superposition between ground and Rydberg states.
From there, rf fields can couple this superposition to additional Rydberg states, shift the EIT signal, and be readout on the amplitude or phase of a probe laser.
Resonant couplings results in Autler-Townes (AT) splittings and can lead to relatively high sensitivities, while off-resonant couplings can be detected through light shifts to the EIT signal and generally result in lower sensitivities \cite{Meyer2020}.
In addition, although the dipole moment of Rydberg atoms scales as the square of the principal quantum number $n$, states with $n \geq 80$ are rarely used as sensors. This is due to the need for increasingly large amounts of laser power to excite such Rydberg transitions, inhomogeneous broadening caused by optical light shifts, and collisional broadening as the Rydberg-Rydberg cross section increases \cite{Fan2015}. All these factors affect the standard EIT sensor configuration employing two optical photons \cite{Sedlacek2012}. It is advantageous, therefore, to develop resonant excitation schemes at moderate $n$ that incorporate transition frequencies relevant to sensing applications. 

Rather than using high $n$ with two-photon EIT, adding additional ``dressing" fields can allow access to higher $L$ angular momentum states which have lower resonant frequencies at moderate $n$.
This can be done with additional optical fields \cite{Brown2022}, or additional microwave fields addressing transitions between Rydberg states \cite{Jin2022, Liu2022}.
Working in the microwave regime has a number of advantages.
From a practical perspective microwave sources are a mature technology that benefit from over a century of investment and development.
By contrast, scientific lasers are typically fragile devices that require extensive external frequency and amplitude stabilization, and are restricted in their tunability.
When using microwave fields, atoms populate multiple Rydberg states \cite{Jin2022}.
Thus, with some adjustments, the system could possibly allow simultaneous sensing using Rydberg transitions resonant with any of the populated states, expanding the available frequencies. 
Additionally, the large number of microwave transitions between Rydberg levels also allows more flexibility to add additional microwave fields and address states with $L>4$ \cite{Facon2016}, or tune Rydberg resonances and extend the sensing range\cite{Berweger2023}.

In this work, we demonstrate a resonant dual-optical, dual-microwave (2O2M) sensor configuration for detecting signals with frequencies near 2.3\,GHz using an $L = 3 \rightarrow 4$ Rydberg transition in Rb.
We discuss the applicability of this scheme across the UHF band utilizing both Rb and Cs, and we model and experimentally compare the sensitivity of this scheme to a more traditional far off-resonant measurement scheme.
Finally, we use the 2O2M scheme to measure SiriusXM satellite radio rebroadcast from a waveguide coupled to the atoms.

In order to detect satellite radio frequencies with our Rb vapor cell based sensor we must be sensitive to 2.3~GHz, near the top of the UHF band.
Resonantly working with frequencies below 2.5~GHz would require the typical EIT scheme to strongly couple to Rydberg levels with $n$ above 90 in Rb, which is generally impractical as mentioned above.
Lower frequencies can be achieved with Cs, but applications below $1$~GHz would still require pushing to $n > 88$.

\begin{figure}
\centering
\includegraphics[width=0.48\textwidth]{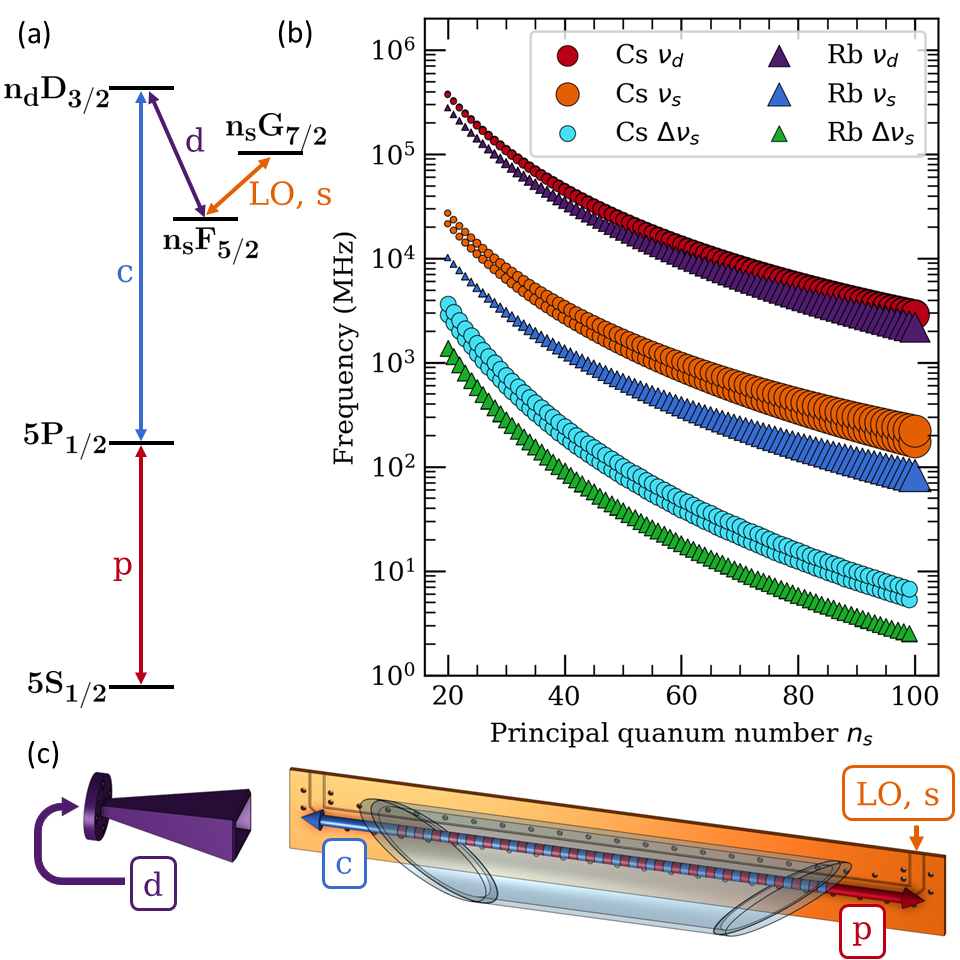}
\caption{(a) Level diagram for the sensor, restricted to the case of $^{85}$Rb operating on the D1 line.
In this work we used $n_d = 35$, and $n_s = 33$.
The probe, coupling, dressing, and combined LO and signal fields are labeled p, c, d, LO, and s respectively.
(b) Frequencies of the $n_s$F$_{5/2, 7/2}\rightarrow n_s$G$_{7/2, 9/2}$ signal transitions in Rb and Cs ($\nu_s$), the corresponding minimum dressing frequency required to reach them ($\nu_d$), and the separation between adjacent signal transitions ($\Delta\nu_s$).
The size of the points are scaled by the matrix element of the transition which range from 290 - 7440 $ea_0$ for the signal transitions.
Operating with a probe laser on the D1 or D2 transition yields two possible signal transitions per $n_s$, which are visibly split on the plot for Cs.
(c) A diagram of the vapor cell and optical and rf fields.
The probe and coupling lasers counterpropagate the cell, the dressing field is produced by a horn colinear with the probe and the signal and LO are provided by a waveguide behind the cell.}
\label{fig:FrequencyVSn}
\end{figure}

Rather than pushing higher in $n$, coupling to higher $L$ states can reach these lower frequencies at much lower $n$.
Fig.~\ref{fig:FrequencyVSn}(b) shows the frequencies accessible with $n_s$F$\rightarrow n_s$G ($L = 3 \rightarrow 4$) signal transitions for a principal quantum number $n_s$ from 20-100 in Cs and Rb.
These frequencies were calculated using the Alkali.ne Rydberg Calculator (ARC) \cite{Sibalic2017}, which is used throughout to calculate resonance frequencies and dipole matrix elements.
In Rb, these transitions can reach down to 300~MHz at an easily accessible $n_s$ of 60, and thus span the UHF band.
This band includes cellular, WiFi, Bluetooth, GPS, and satellite radio frequencies, among others.
In particular, to detect satellite radio with this scheme we use $n_s = 33$.  
In Cs, reaching a given frequency this way requires a higher $n_s$, but in turn will yield a larger dipole matrix element for the sensing transition.
The separation between adjacent signal transitions, also shown in Fig.~\ref{fig:FrequencyVSn}(b), allows denser coverage at lower frequencies. 

\begin{figure}
\centering
\includegraphics[width=.48\textwidth]{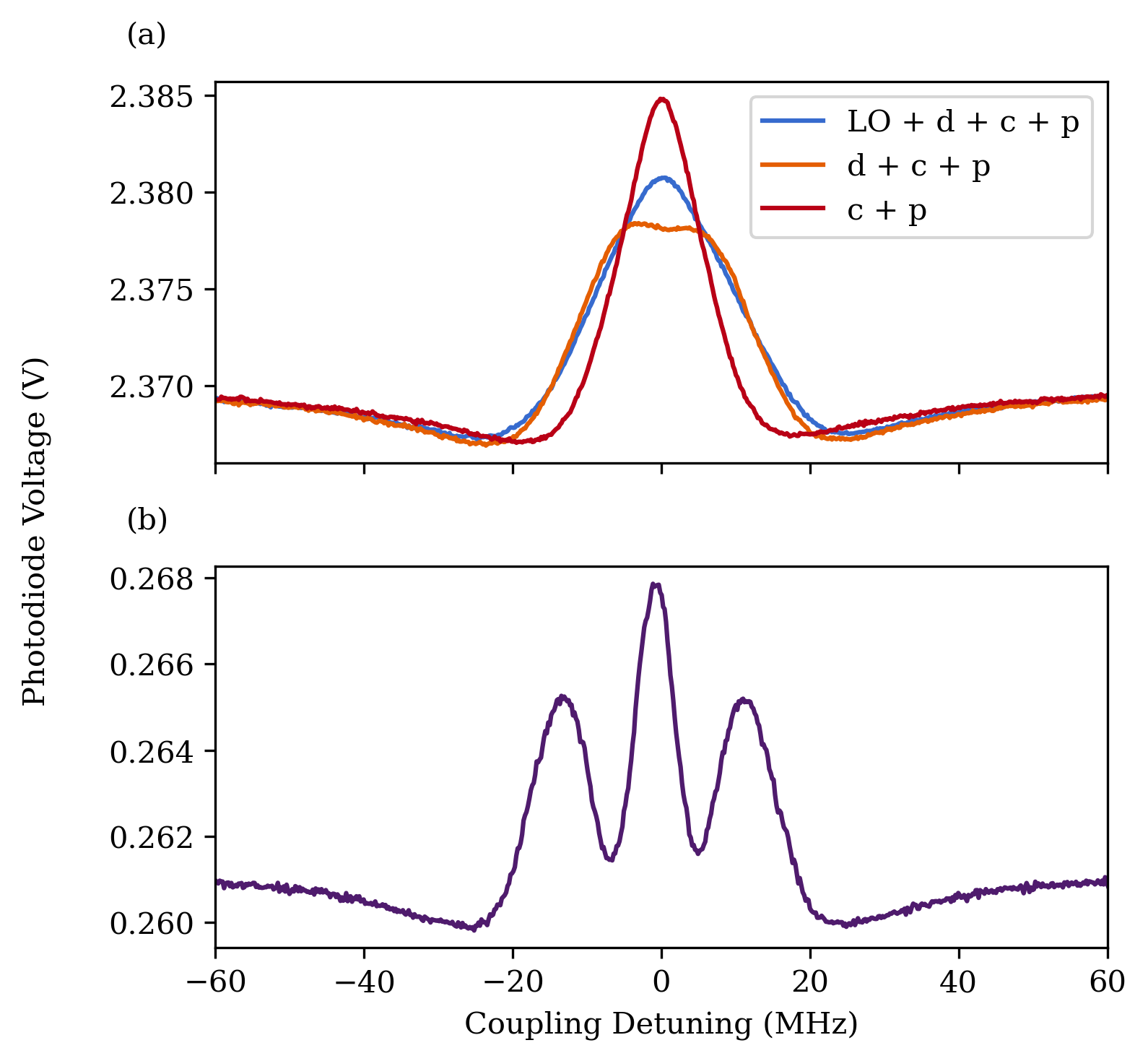}
\caption{Changes in the EIT lineshape from the dressing and LO microwave fields.
(a) The EIT lineshapes at our optimal on-resonance sensitivity parameters (see text) with the LO and dressing fields on (blue), with just the LO on (orange), and with no microwave fields (red).
(b) The full EIT lineshape with a more clearly resolved three peak structure.
This was achieved by increasing the microwave fields for a larger AT splitting and larger central peak, and lower probe power to decrease the linewidths.}
\label{fig:Lineshape}
\end{figure}

To reach an $L = 3$ Rydberg state at least three photons must be used.
In our scheme, corresponding to the level diagram in Fig. \ref{fig:FrequencyVSn}(a), the probe (p) and coupling (c) optical fields reach a Rydberg $D$ state and generate an EIT signal.
The third field, a microwave dressing field (d), operating on a Rydberg-Rydberg transition, addresses an $L = 3$ Rydberg state.
Fig.~\ref{fig:FrequencyVSn}(b) also shows the minimum frequency of this dressing field required to reach the signal transitions.
Once coupled to the $n_sF$ Rydberg state, the atoms are sensitive to the signal field (s) near resonance with the target $L = 3\rightarrow 4$ transition.
We sense the signal field by applying a local oscillator (LO) at the signal frequency and performing rf heterodyne detection \cite{Holloway2019,Jing2020}.
An experimental diagram in Fig \ref{fig:FrequencyVSn}(c) shows how all these fields are supplied to the atoms.
The impact of the additional resonant microwave fields on the EIT lineshape can be seen in Fig.~\ref{fig:Lineshape}.
The dressing field alone creates an AT splitting of the EIT peak, while the LO and signal fields create an additional transmission peak on resonance.
This lineshape is described in detail in Ref. \onlinecite{Jin2022}. The amplitude of the central transmission peak is sensitive to signal fields near resonance, and thus the presence of a signal can be read out in the probe transmission.
Adding further resonant microwave fields alternates between creating transmission peaks and dips on resonance. In principle, this could be used to reach even higher $L$ states and lower frequencies.

Using a semi-classical master equation numerical analysis we now compare the effectiveness of a near-resonant sensor using the 2O2M scheme to a far off-resonant ($\delta_{LO} \gg \Omega_{LO}$) sensor where the dressing field is not present and the signal is sensed through light shifts of the $n_d$D state. 
\cite{rydiqule}.
To do this comparison we define the response of the sensor to the signal field peak amplitude: $d\eta/dE_s$, where $\eta = \text{Im}(\rho_{10})$ comes from the Doppler averaged density matrix and is proportional to the absorption of the probe beam.
This response is inversely proportional to the noise equivalent field (NEF) assuming the sensor is shot noise limited and the probe power is constant.

For the far off-resonant case we can write the response as
\begin{equation}
\frac{d\eta}{dE_s} = E_{LO}\alpha(n)\frac{d\eta}{d\nu},
\end{equation}
where $\alpha(n)$ is the polarizability of the Rydberg state at the signal frequency, and $\nu$ is the center frequency of the EIT window shifted by the LO and signal fields.
We assume that the EIT lineshape and amplitude are independent of $E_{LO}$.
For this case, we modeled a three level EIT system with fixed probe and coupling Rabi frequencies matching our experiment ($2\pi\times6$~MHz and $2\pi\times7$~MHz respectively).
We kept the probe on resonance and optimized over the coupling detuning to find an optimal $d\eta/d\nu = 7.6\times10^{-4}$~MHz$^{-1}$, which is independent of the specific Rydberg state, signal frequency, and LO field amplitude.

For the 202M case we can instead write the response as
\begin{equation}
\frac{d\eta}{dE_s} = \frac{\mu}{\hbar}\frac{d\eta}{d\Omega_s}
\end{equation}
where $\mu$ is the dipole matrix element of the signal transition, and $d\eta/d\Omega_s$ is independent of the chosen Rydberg state.
For this case we modeled the full five-state system with co-aligned linear polarization of all fields.
We maintain the same probe and coupling Rabi frequencies as above, fix the probe laser to be on resonance, and treat the signal as a perturbation of $\Omega_{LO}$.
Then we optimize over $\Omega_d$, $\Omega_{LO}$, $\delta_d$, and $\delta_{c}$.
First we fix the LO on resonance and find an optimal $d\eta/d\Omega_s = -9.2\times10^{-5}$~(Mrad/s)$^{-1}$. 
We then investigate the optimal response of the 2O2M sensor as we detune the signal and LO off-resonance.
We empirically find that once we pass a threshold LO detuning the optimal response is achieved with three robust conditions that are independent of the particular bare EIT signal, or equivalently independent of $\Omega_p$ and $\Omega_c$ in our simulation.
These are that we are on three photon resonance, $\delta_c + \delta_d + \delta_{LO} = 0$, that the coupling laser is resonant with the further detuned Autler Townes peak due to the dressing field, $\delta_c = -\left(\delta_d + \sqrt{(\delta_d^2 + \Omega_d^2)}\right)/2$, and that $\Omega_{d}$, $\delta_d$ and $\delta_c$ scale linearly with $\delta_{LO}$.
Other aspects of the optimization are dependent on $\Omega_p$ and $\Omega_c$.
These include the $\delta_{LO}$ threshold to reach this detuned regime ($\delta_{LO} > 2\pi \times 1.5$~MHz), the generally small $\Omega_{LO}$ relative to the other Rabi frequencies ($\Omega_{LO} \approx 2\pi\times0.1$~MHz), and the particular proportionality constant of $\Omega_{d}$, $\delta_d$ and $\delta_c$ with $\delta_{LO}$ ($\Omega_{d} \approx 8\cdot\delta_{LO}$). 
We have listed the specific scaling of $\Omega_d$ with $\delta_{LO}$ above, which, along with the robust conditions listed above, fully constrains $\Omega_d$, $\delta_d$ and $\delta_c$.
With these conditions met, the response of the sensor is $-1.5\times 10^{-4}$~(Mrad/s)$^{-1}$ and constant with $\delta_{LO}$.
This is an improvement over the on-resonance case, but in practice the necessary $\Omega_d$ will be limited by experimental concerns like the homogeneity of the field and sublevel splittings which are not modeled here, in which case a different local optimum can be found as in our demonstration.

Using the values for $d\eta/d\nu$ and $d\eta/d\Omega_s$ found above  we can calculate the responses for the specific Rydberg states used in our experiment: $n_d = 35$ for the far off-resonant case, and $n_s = 33$ for the transition used in the 2O2M measurement.
For the 2O2M measurement in the detuned regime we get a response of $d\eta/dE_s = -0.01$~(V/m)$^{-1}$, and for the far off-resonant case,  using the polarizability of the $m_J = 1/2$ state, we get $d\eta/dE_s = -2.8\times10^{-8}$~(V/m)$^{-2}\cdot E_{LO}$.
In this case to reach the 2O2M sensor response with the far off-resonant scheme would require an LO field amplitude of $0.4$~MV/m, however this field breaks many assumptions we have made in our analysis.

In our sensor the optical and microwave fields couple the chain of states: 5S$_{1/2} \rightarrow $ 5P$_{1/2} \rightarrow $ 35D$_{3/2} \rightarrow$ 33F$_{5/2} \rightarrow $ 33G$_{7/2}$ in $^{85}$Rb.
This level diagram is shown in Fig.~\ref{fig:FrequencyVSn}(a).
The first two transitions are made by optical probe and coupling lasers with wavelengths of 795~nm and 475~nm respectively.
The second two transitions are made by the microwave dressing field at 119~GHz, and the combination of LO and signal fields at 2.3~GHz each.
Alternatively, one could go through 34D$_{3/2}$ to work with a lower frequency dressing field at 62~GHz, however we were limited by laser power at the required coupling frequency. 

A diagram of the experimental setup is shown in Fig. \ref{fig:FrequencyVSn}(c).
Our experiment was performed in a room temperature vapor cell with natural abundance Rb.
The cell was 12~cm long and each end was at Brewster's angle to minimize reflections.
The probe and coupling beams counterpropagated through the cell forming the interaction region in their overlap.
Both optical beams were vertically polarized, had $1/e^2$ radii of 0.4~mm, and Rabi frequencies $\Omega_p = 6$~MHz, and $\Omega_c = 7$MHz.
After passing through the cell, the probe beam was detected by an amplified photodiode with a bandwidth of 1~MHz which acted as the output of our sensor.

A horn 33~cm away from the cell, and emitting colinear with the probe beam, provided the vertically polarized dressing field, while the LO and signal fields were supplied by a $50~\Omega$ coplanar waveguide 3~mm from the optical interaction region.
The waveguide had a trace width of 3.65~mm, a 0.5~mm gap, and forward voltage gain of $S_{21} = -3$~dB.
Coupling to the waveguide allowed for field confinement and higher sensitivities at a given input power when compared to a horn.
To convert from input power on the waveguide to the field at the atoms we looked at light shifts of the 35D$_{3/2}$, $|m_J| = 1/2$ states \footnote{With optimal positioning of the waveguide only the signal from the $|m_J| = 1/2$ states was present, possibly a pumping effect from the waveguide polarization. This effect allowed us to increase our LO field beyond what is typically optimal in the far off-resonant case.} from the 2.3~GHz field without the dressing field present, and modeled the shifts using a Floquet analysis described in Ref. \onlinecite{Meyer2020}.

Finally, we detected the signal field, whether from a generator or a real world source, by detuning the LO below the desired frequency and measuring the rf heterodyne signal in the transmission of the probe beam.

\begin{figure*}[t]
    \centering
    \includegraphics[width = 7in]{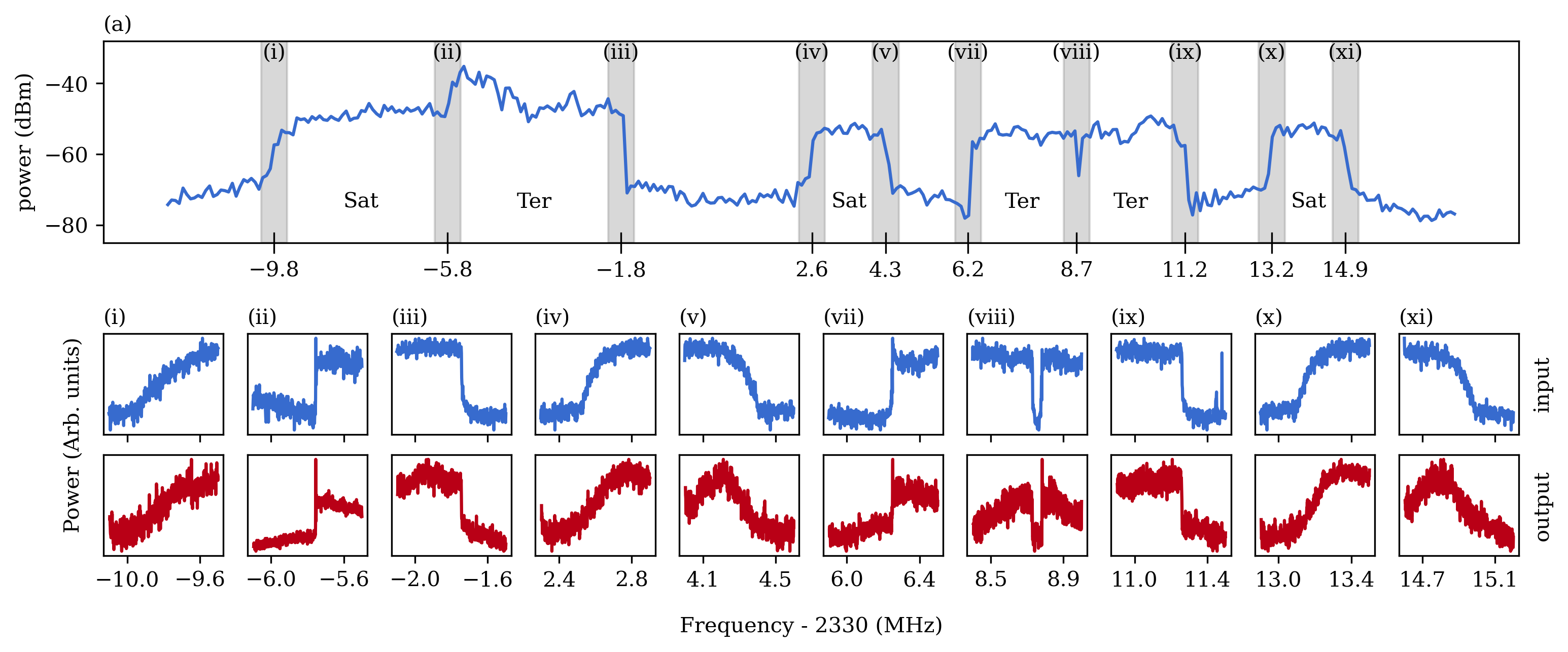}
    \caption{Comparison of the Sirius XM spectrum measured on the input to the waveguide to the output the atomic sensor.
(a) The full spectrum on the input of the waveguide, measured with a 30~kHz resolution bandwidth.
Each band is labeled with its source, either from a satellite directly (Sat) or from a terrestrial repeater (Ter).
(i)-(xi) Zoomed in portions of the spectrum taken at 1~kHz bandwidth.
The input to the waveguide is on top in blue and the output signal from the atomic sensor is below in red.
Each plot corresponds to the edge between two bands and a shaded region in (a).
Some constant spurs have been removed from the atomic sensor output, and the plots have been individually scaled and shifted on a log scale to account for differences from the mirroring effect described in the text.
}
    \label{fig:spectrum}
\end{figure*}

We measured the noise equivalent field of our sensor, both on resonance and detuned from resonance using the 2O2M scheme, and compared it to traditional far off-resonant detection.
Here we define the NEF referenced to the peak field amplitude rather than the root mean squared field amplitude.
These measurements were done by setting the input signal power to a low value where the response of the sensor was confirmed to be linear, then taking the ratio of the power in the heterodyne beatnote to the noise power spectral density at the output of the sensor, extrapolating to find the input power where this ratio would be 1 in a 1~Hz bandwidth, and finally converting from power to field as described previously.

First, we measured the NEF of the on-resonance 202M system to be 70(7)~$\mu$Vm$^{-1}\text{Hz}^{-1/2}$.
For this measurement, the signal field was on resonance at 2.281~GHz, with a peak field amplitude of 44(4)~mV/m corresponding to an input power to the waveguide of -55~dBm.
The dressing field had a Rabi frequency $\Omega_d = 2\pi \times 12$~MHz, and was on resonance at 119.016~GHz.
The LO was set to 300~kHz below the signal field, and had a field amplitude of 1.5(1)~V/m corresponding to an input power of -24~dBm.

To demonstrate the sensitivity away from resonance and prepare for our measurements of satellite radio  
we detuned the signal field by 58~MHz to 2.339~GHz, the center of the XM portion of the Sirius XM spectrum.
To optimize the sensitivity at this frequency we increased the LO field to 4.3(3)~V/m (-15.2 dBm), while maintaining the frequency at 300~kHz below the signal field.
The dressing field was optimal at a Rabi frequency of $\Omega_d = 2\pi \times9$~MHz, and was kept on resonance, while the probe laser was detuned 2.7~MHz above resonance.
These are likely not the globally optimal values, as the numerically found optimum is inaccessible due to the required Rabi frequencies.
With this optimization we found a NEF of 150(10)~$\mu$Vm$^{-1}\text{Hz}^{-1/2}$.

Then we compared the sensitivity from the 2O2M scheme to the more traditional far off-resonant detection at 2.339~GHz in the absence of the dressing field.
In the far off-resonant case the sensitivity can generally be improved by increasing the LO field, so for a fair comparison we amplified our LO up to 520(40)~V/m (26.5~dBm) where we no longer saw improvement in the sensitivity.
In practice, it is often not practical to use such a large LO field, especially without a waveguide or similar structure to provide field confinement.
With this far off-resonant scheme we found a NEF of 4.0(3)~mVm$^{-1}\text{Hz}^{-1/2}$.
This is a factor of 57 (35~dB in power) worse than the 2O2M resonant sensitivity and a factor of 27 (29~dB in power) worse than the detuned 202M sensitivity.

Finally, to demonstrate the viability of this scheme we measured Sirius XM satellite radio from $2.320 - 2.345$~GHz \cite{DiPierro2010} using the experimental parameters for the detuned 2O2M case.
The satellite signal was received by a 10~dBi antenna, filtered and amplified by 54~dB before traveling though roughly 100~m of LMR-400 cable.
The amplification was done in two stages with a bandpass filter in between to avoid saturating the second amplifier.
The cable and the other connections caused $\sim$20~dB of loss on average between the amplifiers and the waveguide.
In total, the field at the atoms from the waveguide was around 166 times larger than the field received by the antenna.
Fig.~\ref{fig:spectrum} (a) shows the full satellite signal at the input of the waveguide measured on a spectrum analyzer with a resolution bandwidth of 30~kHz.
The Sirius XM spectrum is split into three bands from 2.3200 - 2.3325~GHz and six bands from 2.3325 - 2.3450~GHz \cite{DiPierro2010}.
Our antenna picked up six of the nine bands, three directly from satellites and three from terrestrial repeaters.
To view the Sirius XM spectrum using our atomic sensor we set the LO to $\sim$400~kHz below the start of each band, and view the band edge on the output photodiode signal with a spectrum analyzer at 1~kHz resolution bandwidth.
The LO frequencies used were: 2.3198, 2.3238, 2.3278, 2.3322, 2.3339, 2.3358, 2.3383, 2.3408, 2.3428, and 2.3445~GHz.
Fig.~\ref{fig:spectrum} (i)-(xi) show these scans compared to the input power spectral density.
The spectra from the atomic sensor correspond to a combination of the input spectrum above the LO, and a mirrored spectrum from below the LO since the atomic sensor is sensitive to offset frequencies of both signs.
However, this effect only adds a constant offset since the spectrum below the LO set points is largely constant over the range of our measurements.
Obtaining this spectrum using the 2-photon EIT configuration would have required either at least 29~dB more gain in the receiving path or about 5~W of optical power in the coupling field to work with a resonant n=95 transition.

In conclusion, we show that by adding an additional resonant microwave field the opportunity for resonant detection can extend down to hundreds of MHz, at reasonable $n$ levels, and without the need for additional lasers.
We demonstrate an improvement in the NEF by a factor of 50, from 4.0(3)~mVm$^{-1}\text{Hz}^{-1/2}$ in the far off-resonant regime, which additionally requires a high power LO, to 70(5)~$\mu$Vm$^{-1}\text{Hz}^{-1/2}$ with our resonant 2O2M probing scheme at 2.3~GHz.
In addition, we demonstrate the utility of this technique by detecting Sirius XM signals transmitted from satellites and terrestrially.
Extending this work by adding multiple microwave fields to access even higher $L$ states would allow sensitive detection of further lower frequencies. 
In addition, combining this scheme with sensing off of the intermediate Rydberg state could allow for simultaneous detection of multiple frequencies as in Ref.~\onlinecite{Meyer2023} without the requirement that they all couple off of a single Rydberg state.

\begin{acknowledgments}
The views, opinions and/or findings expressed are those of the authors and should not be interpreted as representing the official views or policies of the Department of Defense or the U.S. Government.
References to commercial devices or services do not constitute an endorsement by the US Government or DEVCOM Army Research Laboratory. They are provided in the interest of completeness and reproducibility.
\end{acknowledgments}

\section*{Author Declarations}

\subsection*{Conflict of Interest}
The authors have no conflicts to disclose.

\section*{Data Availability Statement}

The data that support the findings of this study are available from the corresponding author upon reasonable request.

\nocite{*}
\bibliography{Satellite_radio_detection_via_dual-microwave_Rydberg_spectroscopy}% Produces the bibliography via BibTeX.

\end{document}